\documentclass{aa}
\usepackage{graphicx}
\usepackage{natbib}
\usepackage{txfonts}
\usepackage{color}

\begin{document}
 
   \title{Proper motions of spectrally selected structures in the HH~83  outflow\thanks{This work is  based on observations conducted with the 6m telescope of the Special Astrophysical Observatory of the Russian Academy of Sciences. } \thanks{The channel movie is available at... } \thanks{Reduced datacubes (FITS files) are only available at the CDS via anonymous ftp to...}}   

   \author{T.A. Movsessian
          \inst{1}
          \and
          T.Yu. Magakian
          \inst{1}
          \and          
          A.V. Moiseev 
          \inst{2}
          }          
   \institute{Byurakan Astrophysical Observatory, 0213 Aragatzotn prov.,
              Armenia\\
                \email{tigmov@bao.sci.am,tigmag@sci.am}
          \and
              Special Astrophysical Observatory,
             N.Arkhyz, Karachaevo-Cherkesia, 369167 Russia\\
             \email{moisav@sao.ru}
          }      
              
   \date{Received ...; / Accepted ... }

   \abstract{We continue our program of investigation of the proper motions of  spectrally separated structures in the Herbig-Haro outflows with the aid of Fabry-Perot scanning interferometry. This work mainly focuses on the  physical nature of various structures in the jets. } 
{The aim of the present study is to measure the proper motions of the previously discovered kinematically separated structures in the working surface of the HH\,83 collimated outflow. } 
{We used observations from two epochs separated by 15 years, which were performed on the 6m telescope with Fabry-Perot scanning interferometer. We obtained images corresponding to different radial velocities for the two separate epochs, and used    them to measure proper motions. }
{ In the course of our data analysis, we discovered a counter bow-shock of HH 83 flow with positive radial velocity, which makes this flow a relatively symmetric bipolar system. The second epoch observations confirm that the working surface of the flow is split into two structures with an exceptionally large (250~km\ s$^{-1}$) difference in radial velocity. The proper motions of these structures are almost equal, which suggests that they are physically connected. The asymmetry of the bow shock and the turning of proper motion vectors suggests a collision between the outflow and a dense cloud. The profile of the    H$\alpha$ line for the directly invisible infrared source HH~83~IRS, obtained by integration of the data within the reflection nebula, suggests it to be of  P~Cyg type with a broad absorption component characteristic of the FU~Ori-like objects. If this object underwent an FU~Ori type outburst, which created the HH~83 working surfaces, its eruption took place about 1500 years ago according to the kinematical age of the outflow.}
{} 
   \keywords{ISM: jets and outflows -- ISM: individual objects: HH~83 -- Stars: pre-main sequence -- Stars: individual: HH~83~IRS (IRAS 05311$-$0631)}
   \titlerunning{Proper motions in the HH83 outflow}
   \authorrunning{Movsessian et al.}
     \maketitle
%
\section{Introduction}

High proper motions (PMs) of Herbig-Haro (HH) objects were discovered about 40 years ago.  The orientation of the PM vectors of HH objects, which represent  shocked excitation regions,  revealed the bipolar nature of high velocity flows responsible for their formation \citep{Herbig&Jones1981,Jones&Herbig1982}. Further discoveries of  highly collimated jets from young stellar objects indicated that HH objects are the brightest parts of these flows \citep{Reipurth1986,Reipurth1993}; in subsequent studies these were named HH jets and flows. 

Further investigations of HH flows revealed their complex morphology. In particular, high- and low-excitation zones, which were separated by ground-based and Hubble Space Telescope (HST) narrow band imagery \citep{Reipurth1997, Hartigan2011} as well as by long-slit spectroscopy \citep{Heathcote1992, Reipurth1997, Hartigan2011}, were revealed in terminal working surfaces (WSs) and inside certain knots in the  jets. The full morphology  and kinematics of these structures can be understood using methods of spectral imaging \citep{Hartigan2000,Mov2000,mov2009}.

Briefly, inside the terminal WSs,    which is the common term for the regions where supersonic flow slams directly into the undisturbed ambient medium  \citep{Reipurth2001}, as well as in the internal WSs of jets, two principal shock structures form: a
``reverse shock'', which decelerates the supersonic flow, and a ``forward shock'',
which accelerates the ambient material with which the flow collides \citep[e.g.][]{hartigan1989}.

In addition,  various other shocked structures, such as deflection shocks, evacuated cavities,  intersecting shock waves, Mach stems, clumps, and sheets, can form in HH flows \citep{Hartigan2011}. The study of such structures  can help us to obtain a global understanding of the processes taking place inside the collimated outflows. Proper motion
measurements of already separated kinematical structures are of great importance for the discovery and analysis of all the above types of shocked structures. 

The subject of the present work is the \object{HH~83} system. Its wiggling knotty jet emerges from \object{IRAS 05311$-$0631} source. This source, being deeply embedded in a molecular cloud, is not visible in the optical range, but it illuminates the edges of a conical cavity formed by the outflow \citep{reipurth89}. At the distance of 105\arcsec\ from the beginning of the jet, a bow shock structure was found, 
which is visible in H$\alpha$ emission, but not in [\ion{S}{ii}] lines \citep{reipurth89}. 

Our previous observations of HH~83 with scanning Fabry-Perot interferometer (FPI) revealed separated structures with very different radial velocities in the terminal WS (more than 250 km\ s$^{-1}$), and confirmed the steady increase in velocity along the jet depending on the distance from the source \citep{mov2009}. Furthermore, unusual wave-like variations in the radial velocity of the jet  as a function of distance from the source are also visible in the values of full width at half maximum (FWHM) of emission lines.

 The present investigation was motivated by our    two-epoch study of the \object{HL~Tau} jet, where observations from 2001 and 2007 were used to estimate the PM of spectrally separated structures in the jet.
This method, applied to HH jets for the first time to the best of our knowledge,  reveals that the structural
components inside the jet, which can be divided into components of low- and high-radial velocity, have, nevertheless, very
similar values of PM \citep{Mov2012}. This important result
was also  confirmed for the 
components inside the \object{FS~Tau~B} jet system \citep{mov2019}. The HH~83 system  represents the next suitable object for second epoch spectra-imagery observations, especially in view of the discovery of two separated kinematical structures in its WS. However, HH~83 is a much more distant object than the Taurus dark cloud. Taking this into account we decided to increase the time between the two epochs to up to 15 years.

\section{Observations and data reduction}

Observations were carried out in the prime focus of the
6m telescope of the Special Astrophysical Observatory of the Russian Academy of Sciences in two epoches: 10 February 2002 and
2 February 2017, in good atmospheric conditions (seeing was about
1\arcsec). We used a FPI placed in the collimated beam of the Spectral Camera with Optical Reducer for Photometric and Interferometric Observation (SCORPIO) \citep{AM2005} and SCORPIO-2 \citep{AM2011}  multi-mode focal reducers in 2002 and 2017, respectively. The
capabilities of these devices in scanning FPI observational mode are presented in \citet{moisav, moisav2015}. 
A description of the first epoch observations was presented by \citet{mov2009}. 

During the second epoch of observations, the detector was a EEV 40-90 $2\times4.5$K  CCD array. The observations were performed with 4$\times$4 pixel binning, and so 512 $\times$ 512 pixel images were obtained  for each spectral channel. The field of view was 6.1\arcmin\ with a scale of 0.72\arcsec\ per pixel. The second epoch of observations provided deeper images and of higher spectral resolution. 

The scanning interferometer was ICOS FPI operating in the 751st order of interference at the H$\alpha$ wavelength, providing a spectral resolution of FWHM $\approx$ 0.4\AA\ (or $\approx$20 km s$^{-1}$) for a range of $\Delta\lambda$=8.7\AA\ (or $\approx$390 km s$^{-1}$) free from order overlapping. The number of spectral channels was 40 and the size of a single channel was $\Delta\lambda\approx$ 0.22\AA\ ($\approx$10 km\ s$^{-1}$).
In both epochs an interference filter with FWHM $\approx$ 15\AA\ centered on the H$\alpha$ line was used for pre-monochromatization.
 
We reduced our interferometric observations using the software developed at the SAO \citep{moisav,moisav2008,moisav2015} and the ADHOC software package{\footnote{The ADHOC software package was developed by J. Boulestex (Marseilles Observatory) and is publicly available on the internet.}}. After primary data reduction, subtraction of night-sky lines, and wavelength calibration, the observational material was organised into ``data cubes''. We applied optimal data filtering, which included Gaussian smoothing over the spectral coordinate with FWHM = 1.5 channels and spatial smoothing using a two-dimensional Gaussian with FWHM = 2--3 pixels.
The FPIs used in the two epochs of observations had different spectral resolutions. Therefore, the rebinned data cubes were created for both epochs to bring them into the same velocity steps. This allows a better comparison of morphological details that have the same radial velocities.
All radial velocities presented in this paper are heliocentric.
 
Using these data cubes, we spectrally separated  the
details with a different radial velocity in the outflow system. Proper motions  were then measured for  the selected structures using observations in both epochs. For the PM estimation, we used a
method of optimal offset   computation for two images by
means of cross-correlation (this is the `IDL' procedure by F. Varosi and included in the IDL astronomy library\footnote{\textit{https://idlastro.gsfc.nasa.gov/}}).

\section{Results }

 During the observations, the field of view of the SCORPIO focal reducer covered  the entire HH~83 outflow system including the jet, the WS, and the reflection nebula around the source 
(which actually represents the illuminated cavity walls). Below we discuss all these parts of HH~83 separately and compare the observations of 2002 and 2017.

\begin{figure}
\centerline{\includegraphics[width=20pc]{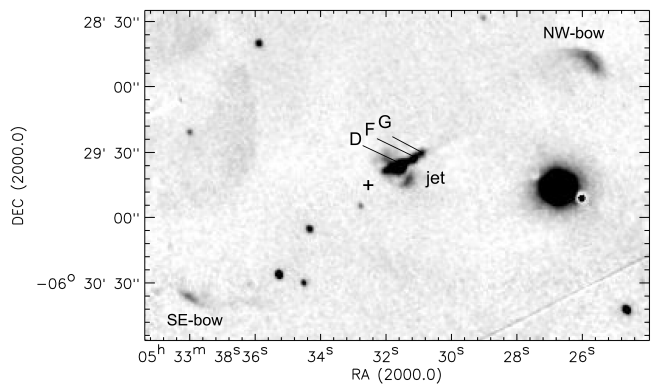}}
\caption{ Monochromatic image in H$\alpha$ line of the HH~83 outflow system, which consists of a narrow jet (main knots are marked) and two WSs in NW and SE directions from the source (seen only in the infrared range  and marked here by a cross).  } 
\label{field}
\end{figure}

Due to the higher quantum efficiency (QE) of both the new detector and the whole optical system, the data from the  second epoch have a better S/N in comparison with those of the first epoch. Figure\,\ref{field} shows the integrated H$\alpha$ image  of the HH~83 system built using the second epoch observations. The main details are marked.

\subsection{The terminal working surfaces, their proper motions, and kinematical structures}

Besides the already known WS (further denoted as NW-bow) which lies at a distance of about 2\arcmin\  from the source, our image clearly shows a second bow-shape structure in the opposite direction from the source, not described previously (Fig.\,\ref{field}). This structure lies on the axis of the outflow system  at nearly the same distance from the source as the main NW-bow. After re-examining the first epoch data, we found faint traces of this second
bow-shaped structure in the same place.  In contrast with the NW-bow, this second bow shock has positive radial velocity. It undoubtedly represents the terminal WS of the counter flow (further SE-bow), confirming the bipolar and symmetric nature of this outflow system. 

\begin{figure*}
\centerline{\includegraphics[width=30pc]{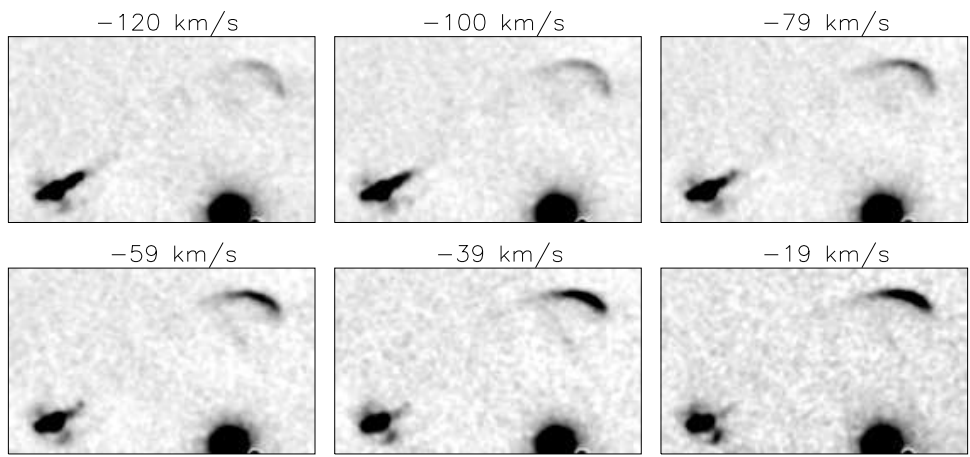}}
\caption{Velocity channel map of HH~83 outflow system showing morphology variations of the low velocity structure in NW bow}
\label{chmap}
\end{figure*}

In
the first paper \citep{mov2009} we mentioned the significant asymmetry of the NW-bow as well as the trend of the decreasing  radial velocity from the apex toward the end of the wing of the bow shape structure. Using
new observations, we built a velocity channel map of low-velocity structures in the NW-bow of the HH~83
outflow system (Fig.\,\ref{chmap}). This channel map confirms that the low-velocity component of the NW-bow shows
noticeable differences in morphology depending on the radial velocity. At relatively high velocities, it has a more or
less symmetric shape, and at lower velocities it becomes more and more asymmetric. 

As mentioned above, our previous data demonstrated that the NW-bow in the HH~83 system is divided into two 
distinct components, which are spectrally and spatially  well separated \citep[see Fig. 6 in][]{mov2009}. After Gaussian fitting of both components of H$\alpha$ emission we restored images of these components for each  observational epoch.  Using these images we measured the PMs of high- and low-radial-velocity structures in the NW-bow. 

\begin{figure*}
\centerline{\includegraphics[width=36pc]{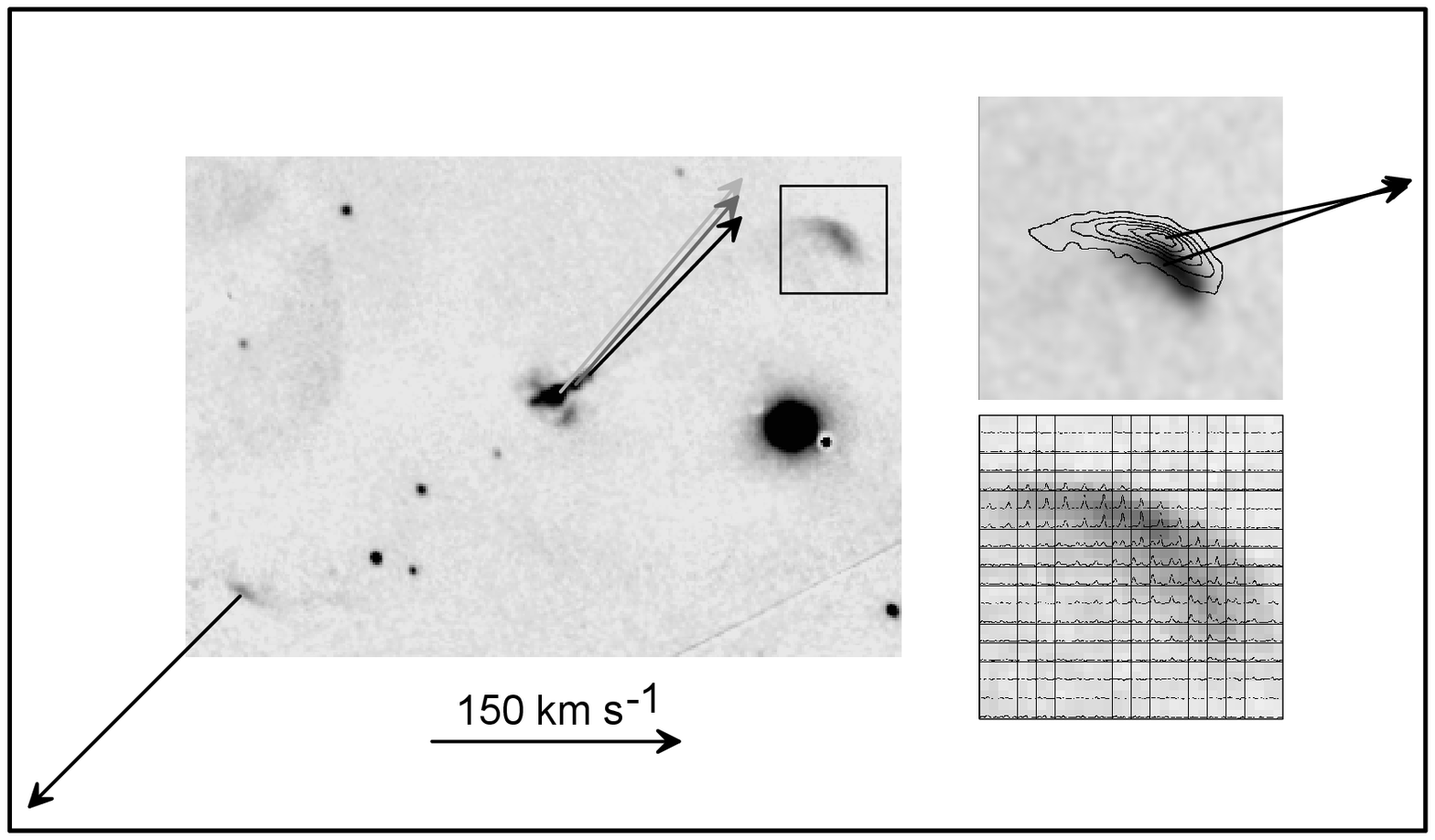}}
\caption{H$\alpha$ image of the HH 83 outflow system integrated over all velocity channels,  with superposed PM vectors for the jet 
knots and the counterflow (left panel). The bottom right panel shows the grid 
of H$\alpha$ profiles in the NW bow. The 
significant split of profiles for the high- and low-velocity components is obvious (the absolute velocity values decrease to the right).  Separate images of 
the low-velocity (contours) and high-velocity components (grey scale) based on the intensities obtained by the Gaussian fitting of the profiles  of both
components  are presented in the  top right panel. The corresponding PM 
vectors also are shown. }
\label{PMall}
\end{figure*}

\begin{table*}
\caption{Proper motions and radial velocities of knots in the HH 83 outflow}
\label{table}
\centering
\begin{tabular}{l c c c c c}
\hline\hline
Knot & Distance & V$_{tan}$\tablefootmark{a}  & PA & V$_{r}$ &  V$_{abs}$\\  & (arcsec) & (km
s$^{-1}$) &(deg)
&   (km s $^{-1}$) &  (km s $^{-1}$) \ \ \ \ \\
\hline
D               &  13.5 &   120  $\pm$ 30  &   310    &   $-130$  &177\\
F               &  19.3 &   142  $\pm$ 35  &     327    &   $-140$ & 200\\
G               &  27.8 &   135 $\pm$ 40  &     319  & $-165$ & 213\\

WS $_{high~vel}$  &  117.0   &   160 $\pm$ 23 & 282  & $-312$ & 351\\
WS $_{low~vel}$   &  119.4   &   145 $\pm$ 30         & 277  & $-50$ & 153\tablefootmark{b}\\
counter bow       &   96     &   180 $\pm$ 45         & 135    & +180  &  255   \\
\hline
\newline
\end{tabular}
\tablefoot{
\tablefoottext{a} {Tangential velocities correspond to a distance of 450 pc for the flow}
\tablefoottext{b} {The value of this velocity is discussed below in Sect. 4.1}
}
\end{table*}

The results of our PM measurements are presented in Fig.\,\ref{PMall}.
In particular, the PM vectors for both the low-
and high-radial-velocity structures of  the NW-bow of the HH~83 flow are shown in the inset. Both structural
components can be seen to have very
similar values of PM, despite the large difference in radial velocity. We also estimated the distance between these structures during the second epoch observations
and we find this distance to be the same as for the first epoch (1000 AU). This is additional evidence of their equal PM values. We discuss this result in more detail in the following section.

 In Fig.\,\ref{PMall} we also show the PM vectors for the three brightest knots in the HH~83 jet, and for the separated structures in the NW-bow and the SE-bow, that is for the counter bow shock. It is worth mentioning that this second bow shock has nearly the same range of  radial velocities (in absolute values) as the NW-bow, but, unlike the latter, it  does not show either a spectral or a spatial  inner separation. In any case, we were able to measure the PM of this structure,  even though it was significantly fainter in the first epoch data. 

 The numerical results are given in Table\,\ref{table} where we present the distances for each structure measured from the source position,
values of tangential velocities (computed for the distance of 450 pc), position angles (PAs) of PM vectors, radial velocities, and calculated absolute spatial velocities.

\subsection{Proper motions of the knots in the jet}
We measured the PMs of several bright knots in the HH~83 jet, namely knots D, G, and F \citep[according to the nomenclature of][]{reipurth89}.
These knots also show high values of PM (see Table\,\ref{table}) similar to those of high- and
 low-velocity structures in the terminal WS. It should be noted, however, that although the geometric axis of the jet coincides with that of the NW-bow, the position angles of the PM vectors of the jet knots significantly differ from those of the structures in the NW-bow: they are  turned in the northern direction for about 14 degrees. The spatial velocities of the knots, calculated from their radial and tangential velocities, increase along with the distance from the source. The same trend is observed in their radial velocities (see Table\,\ref{table}). We discuss this effect in the following section.

The knots in the HH~83 jet do not demonstrate a distinct division into high- and low-radial-velocity components; consequently  it was impossible to reveal the morphological structures with different radial velocities inside them.

\subsection{Radial velocities of the system in general}

\begin{figure}
\centerline{\includegraphics[width=16pc]{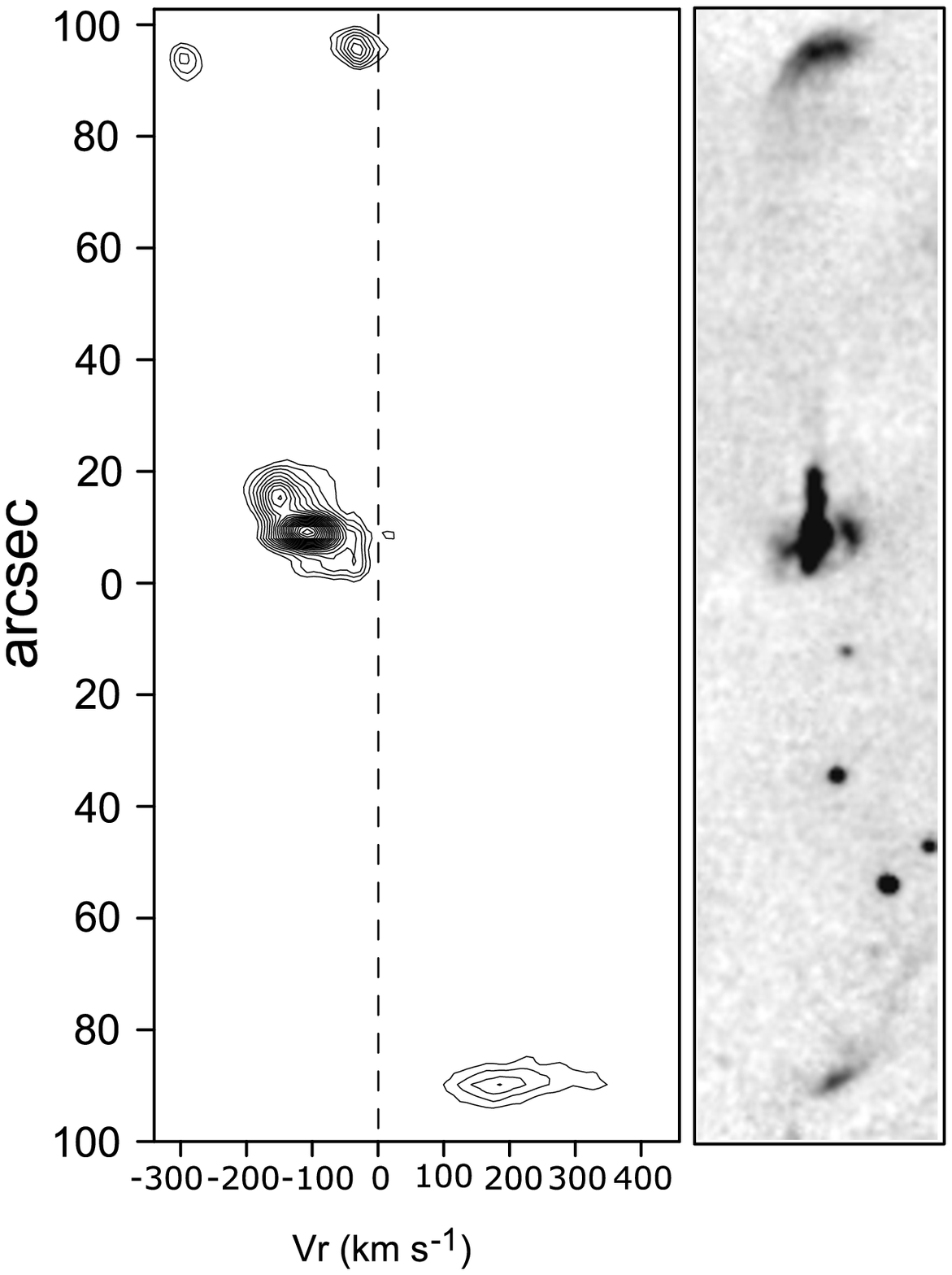}}
\caption{ Position-velocity diagram along the full length of
the HH~83 system, including both WSs and the jet. The NW WS clearly separates into two distinct structures. } 
\label{PV}
\end{figure}

The radial velocities estimated from the second epoch FP observations are in good agreement with those of the first epoch. Figure\,\ref{PV} shows the position velocity diagram and the integrated H$\alpha$  image of the HH~83 outflow. New observations with higher spectral resolution confirm an increase in the negative radial velocity with distance from the source, and also confirm the remarkable split in radial velocity     inside the terminal NW-bow. The counter bow-shock (SE-bow) has high positive velocity
over a significant range (FWHM $\sim$ 150 km s$^{-1}$). The whole  system represents a bipolar collimated flow with a full size of about 0.48 pc.

\begin{figure}
\centerline{\includegraphics[width=12pc]{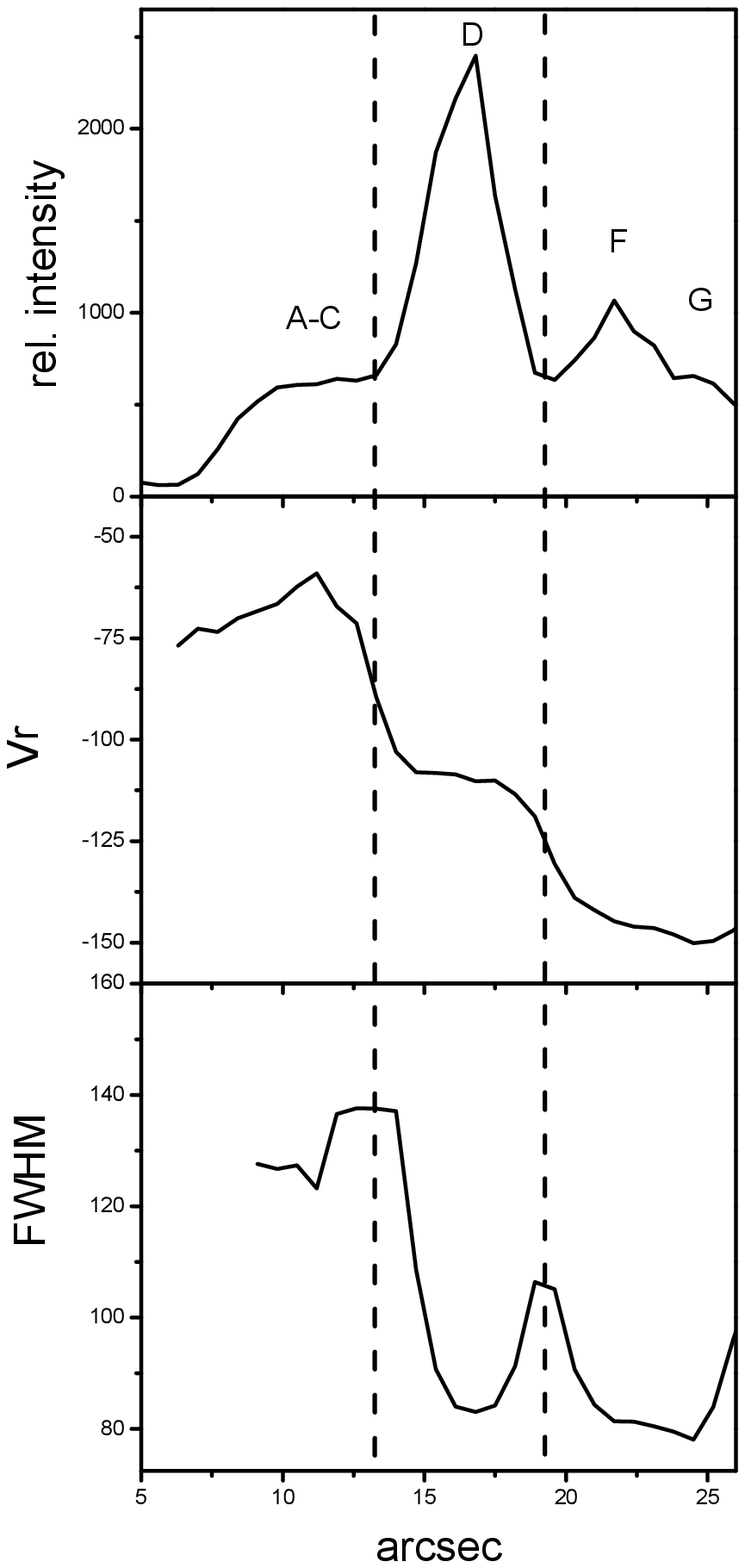}}
\caption{ Plots of the intensity, radial velocity, and FWHM of the  H$\alpha$ emission in the HH~83 jet versus angular distance from the source. The abrupt rise in FWHM after 25\arcsec\ is spurious and caused by low S/N.}  
\label{waves}
\end{figure}

To compare the observations of both epochs, we built the distributions of radial velocities and FWHM of H$\alpha$ emission along the HH~83 jet itself (Fig.\,\ref{waves}). These show a good  match to the previous data shown
in Figs. 3 and 4 of \citet{mov2009} and confirm the existence of wave-like variations in the general trend of these parameters.
We discussed these variations in this latter paper, but the new data provide higher spectral resolution and S/N. The prominent steep drops between knots A and D, as well as between D and F, are clearly seen in Fig.\,\ref{waves}. In these regions the steep increase in FWHM is observed as well; though in general we see a constant
decrease in FWHM values along the jet, as in other cases.

\subsection{The reflection nebula }
\begin{figure}
\centerline{\includegraphics[width=14pc]{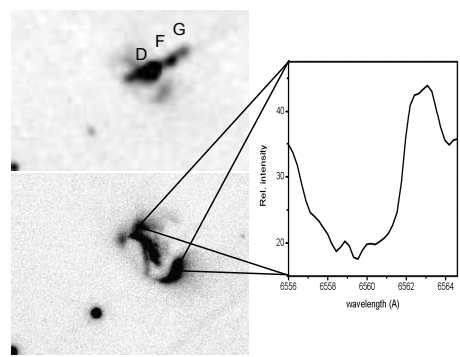}}
\caption{Integrated profile of the H$\alpha$ line in the reflected light of the source of HH~83. The left side shows an enlarged image of the central part of HH~83 system in  H$\alpha$ emission (the integrated FP image from 2017, upper panel) and in \textit{i} band (PanSTARRS survey, lower panel).}
\label{source}
\end{figure}

Observations show that the HH~83 jet is propagating through the evacuated conical cavity formed by wide-angle wind from the deeply embedded infrared source \citep{reipurth2000}. The walls of this cavity are illuminated by light from this source. In the optical range, these walls form two reflection lobes around the narrow jet. In the high-resolution infrared images  \citep{reipurth2000}, several spiral or helical arms can be seen on the cone walls, somewhat similar  to the structures observed in \object{GM~3-12} (\object{RNO~124}) nebula \citep{Mov2004}. On the PanSTARRS \textit{i} band image (see Fig.\,\ref{source}, lower panel) these  two lobes are connected by a bar-like structure, which probably represents one of the above-mentioned arms. On the restored H$\alpha$      image  (Fig.\,\ref{source}, upper panel) these details are almost invisible, which confirms their reflection nature.  

 The above-mentioned high quality of the new data allows an attempt to obtain  the profile of the H$\alpha$ line in the spectrum  of the central star through the scattered light of the source deeply embedded in the dark cloud. To study the profile of the H$\alpha$ line in the reflected light we used spectral imaging, which allows the data to be  summarized and the total line profiles to be constructed for any selected area.  To avoid contamination from the jet radiation and to deal  with pure continual (i.e., reflected) light, we integrated profiles from the two side lobes only, excluding the bar-like structure, which is superimposed on the D knot of the jet. The resulting integrated H$\alpha$ profile is presented in the right panel of  Fig.\,\ref{source}. This profile has strong and wide ($\sim$300 km\ s$^{-1}$) blueshifted absorption with a weak emission component near the zero radial velocity. We assume that this profile corresponds to the invisible IR source.

\section{Discussion}
\subsection{Working surface}

 Comparing the inner structures in the WS of HH~83 and in the knots in the HL~Tau jet, in both cases we find two emission structures with very different radial velocities. However, in the HL~Tau jet these structures  differ not only in their radial velocities
but also in excitation, while in parts of the HH~83 WS the excitation level   does not change, and these structures are visible mainly in H$\alpha$ \citep{reipurth89}. 

The radial velocities of these two emission structures differ by more than 250 km\ s$^{-1}$. This is a unique case: for example, two velocity components in the WS of \object{HH~111}  differ by about 60 km\ s$^{-1}$ only \citep{reipurth97} and the two components  in the HL~Tau jet differ by about 100 km\ s$^{-1}$.  This large difference in speed cannot be explained by the existence of two separate outbursts, which formed two bow-shape structures with strongly different radial velocities, because in this case the high-velocity structure will catch up to the low-velocity one in 15 years; in addition,
our new observations reveal that the PMs of these structures are the same. 

 Therefore, we conclude that we observe physically connected structures  in the terminal WS of HH~83 that can be designated a reverse shock and a forward shock. The observed large difference in the radial velocities of these structures can form in very extreme conditions. For example, in the case of the HL~Tau jet, the bright low-velocity structures in front of the high-velocity knots appear only in the region where the collisional interaction  between the jet and wide outflow from XZ~Tau is taking place \citep{Mov2007}.

The asymmetry of the bow-shock as well as the turning of its PM vectors compared to those of the jet knots are further indications of the encounter between the jet and the dense cloud core. A\ well-known example of a similar jet collision with dense cloud is \object{HH~110} \citep{reipurth96, lopez2010}.

Here we observe the bow with increased brightness of the wing, probably formed by an oblique collision. Analysis of the
morphology of the  bow   in various radial velocities using a velocity channel map (Fig. \ref{chmap}) shows that
the bow is more symmetric at higher velocities, and the apex of the bow can be seen in the channels
corresponding to radial velocities near 100-120 km s$^{-1}$. However, this value is  still much
lower than that for the reverse shock region.

Taking into account the simple bow-shock and Mach disk model of Hartigan (1989), such a velocity difference could occur when the ratio of the density of the ambient medium to  that of the jet is near 100, and with a jet velocity of more than 300 km\ s$^{-1}$.

If our conclusion is correct and the high-velocity structure in the WS indeed represents a reverse shock, then we detect emission of the decelerated flow matter; the ratio of pattern speed to flow speed in this structure will,  by definition, be equal to one. On the contrary, for the forward shock, which represents accelerated matter from the ambient medium, this ratio will be  much higher because of the lower radial velocities of the emitting particles. Based on this, we calculated the spatial velocities of the flow structures, presented in the Table \ref{table}; one should keep in mind that only the values for the reverse shock represent the real flow velocity.   In this way, it is possible to compute the inclination angle  between the line of sight and the jet, which we find to be close to 27 degrees. Using all these values, one can estimate the kinematical
age of the outflow to be about 1500 years.

In general, the structures with different radial velocities in the terminal WS of HH~83 outflow are similar to the inner structures in the HL~Tau jet. But there are certain differences. The high-velocity structure in the WS of HH~83 is not as compact as in the case of the HL~Tau jet and represents an extended object consisting of several condensations.
We think that this is caused  by fragmentation of the jet matter during its collision with a dense
cloud core. Presumably, it forms a common forward shock region, which we see as a narrow bow-shape structure.

\subsection{The jet }
Unfortunately the spatial resolution of our observations is not sufficient to resolve various velocity structures in the knots of the HH~83 jet. As mentioned above, the  data obtained in 2017 confirm the increase in absolute radial velocity values with distance from the source. The obtained PMs of the knots  also indicate a small increase in tangential velocities. Using both velocity components, we calculated the spatial velocities for the three knots; they show the same behavior. This result rules out an explanation of the increase in radial velocities by bending of the jet. The wave-like oscillations superposed on the general increase in velocity can be the result of the quasi-periodical variations in the rate of matter ejection. The FWHM along the jet peaks between the knots. However, a thorough analysis of the profiles shows that this effect (in certain places even a split of emission is detected) could be the result of overlapping neighboring knots caused by seeing effects. On the other hand, high values of FWHM can be seen in the starting part of the jet, near the knots A-C. As these knots are located in the zone where the outflow enters the cavity in the dark cloud, such a large range of velocities could perhaps be related to the interaction of the matter of the outflow with the cavity walls. 

\subsection{The source IRAS 05311$-$0631 }

The analysis of the reflected spectrum of the HH~83 source shows that it has a  P~Cyg-type H$\alpha$ line profile with a wide, almost rectangular absorption and a faint secondary blueshifted peak. Such profiles are likely to
be formed in a strong wind with optical depth sufficient  to produce the deep blueshifted
absorption component \citep[e.g.,][]{muzerolle}; they are highly typical of FU~Ori-type eruptive stars; on the other hand, only a few T Tauri stars show such well-developed P Cygni profiles at H$\alpha$ line. The blue edge of the absorption trough indicates wind velocities of up to 300 km\ s$^{-1}$. This value is nearly the same as the velocity of the powerful wind that emanates from \object{FU Orionis} \citep[e.g.,][]{Herbig2003}.

The detection
of faint CO absorption bands at 2.3 $\mu$m by \citet{davis2011} can be considered as a further argument in favor of the FU~Ori-type nature of IRAS 05311$-$0631. Such absorption bands, though usually more strong, are typical for FUors \citep[e.g.,][]{ra1997} .

On the other hand, FU~Ori phenomena in nearly all cases are
associated with HH outflows \citep{audard}. Moreover, the spacing of the  structural details in some HH outflows corresponds to the statistical estimates of
  the supposed recurrence of the FUor events (Herbig et al.
2003). In this case, the bright WS of the HH~83 flow could be the result of an FU~Ori-type
outburst, which took place about 1500 years ago judging by the kinematical age of the outflow.

\section{Conclusion}
This work is a continuation of a series of investigations of PMs of spectrally separated
structures in collimated outflows from young stars based on observations with a scanning Fabry-Perot
interferometer conducted in different epochs. Taking into account the discussions above  in conjunction with previously
obtained results, we draw several conclusions, and make some further suggestions as outlined below.

\begin{enumerate}
\item During our observations we discovered a second WS in the opposite direction to the
previously known WS, which indicates the bipolar nature of the HH~83 outflow system.

\item
Our estimation of the tangential velocities of the 
two previously discovered kinematically distinct
structures in the NW-bow indicates similar values.

\item
We are inclined to assume that the structures of different velocity  in the NW-bow represent two
principal shocks, namely reverse and forward shocks, as in the case of  the internal WSs of the HL~Tau jet.
\item
The asymmetry of the NW-bow and the very large difference in radial velocities for the two separate structures lead
us to conclude that the flow collides with dense cloud in this region. 
\item
The increase in the spatial velocity of the knots in the HH~83 jet once again shows that the behavior of radial
velocities cannot be explained by jet bending.
\item
Our analysis of the H$\alpha$ line in the reflection nebula allows us to define the nature of the
source of outflow system ---which is deeply embedded in a molecular cloud and is not visible in the optical range--- as an
FU~Ori-type object that underwent outburst about 1500 years ago (taking into account the kinematical age of
the NW WS).
\end{enumerate}

We hope that this work will
be useful for the  development of new theoretical models of collimated outflows and of their interaction with the interstellar medium. The  confirmation of HH~83~IRS (IRAS 05311$-$0631) as an FU~Ori-like object is important because of the very small number  of known representatives of this class of eruptive stars. Further study of this object  will help to gain a better understanding of these phenomena.

\begin{acknowledgements} 
Authors thank referee for the helpful comments. This work was supported by the Science Committee of RA, in the frames of the research project 18T-1C329. Observations with the SAO RAS telescopes are supported by the Ministry of Science and Higher Education of the Russian Federation (including agreement No05.619.21.0016, project ID RFMEFI61919X0016).
The Pan-STARRS1 Surveys (PS1) have been made possible through contributions of the Institute for Astronomy, the University of Hawaii, the Pan-STARRS Project Office, the Max-Planck Society and its participating institutes, the Max Planck Institute for Astronomy, Heidelberg and the Max Planck Institute for Extraterrestrial Physics, Garching, The Johns Hopkins University, Durham University, the University of Edinburgh, Queen's University Belfast, the Harvard-Smithsonian Center for Astrophysics, the Las Cumbres Observatory Global Telescope Network Incorporated, the National Central University of Taiwan, the Space Telescope Science Institute, the National Aeronautics and Space Administration under Grant No. NNX08AR22G issued through the Planetary Science Division of the NASA Science Mission Directorate, the National Science Foundation under Grant No. AST-1238877, the University of Maryland, and Eotvos Lorand University (ELTE).

\end{acknowledgements}


\end{document}